\documentclass{stsci_report}

\usepackage{multirow}
\usepackage{array}
\usepackage{graphicx}
\usepackage{siunitx}
\usepackage{todonotes}
\usepackage{pdfpages}
\usepackage{longtable}
\usepackage{booktabs}
\usepackage{float}
\usepackage{longtable}
\usepackage{indentfirst}
\usepackage{caption}
\usepackage{hyperref}
\usepackage{multirow}
\usepackage{paracol}
\usepackage{colortbl}
\usepackage{setspace} 
\usepackage{amsmath}
\usepackage{enumitem}
\usepackage{soul}
\usepackage{xcolor}
\usepackage{hyperref}
\usepackage{array} 
\hypersetup{
    colorlinks=true,
    linkcolor=black,
    urlcolor=blue,  
    citecolor=black
}
\bibliography{bibliography.bib}

\copyrighttext{Copyright\copyright\ \the\year\ The Association of Universities for Research in Astronomy, Inc. All Rights Reserved.}

\presubtitle{Instrument Science Report WFC3 2026-01}
\title{Updates to the WFC3/IR Photometric Stability Stellar Cluster Study}
\author{K.~Huynh \& V.~Bajaj}
\date{May 6, 2026}

\begin{document}

\maketitle

\abstract{The WFC3/IR channel has been shown to lose sensitivity over time. The sensitivity loss rate has been quantified in past works and we extend those measurements into 2026 using photometry of uncrowded regions of star clusters 47 Tucanae and Messier 4 in the F110W and F160W filters. This study adds 4--7 additional visits since \textcite{2022Bajaj} for each target/filter combination. Using F160W data of 47Tuc, we find that the IR channel continues to lose sensitivity at a rate of  -0.07\%~$\pm$~0.01\% per year using the full dataset, and -0.06\%~$\pm$~0.02\% per year using only the first exposure of every visit. Using M4 data observed in F110W, we find a sensitivity loss rate of  -0.13\%~$\pm$~0.02\% per year using the full dataset, and -0.07\%~$\pm$~0.03\% per year using just the first exposure of every visit. We find that the rate of sensitivity loss has decreased since the previous study. This likely stems from the inclusion of more recent visits that were designed to mitigate persistence effects. These results suggest that earlier observations, which were not intended for sensitivity measurements, may have overestimated the IR channel’s sensitivity due to the effects of persistence from bright sources. The \texttt{IMPHTTAB} reference file that contains the correction to the time-dependent sensitivity remains unchanged from the sensitivity measurements of this report.}

\section*{Introduction}
The Wide Field Camera 3 Infrared  (WFC3/IR) channel is known to exhibit changes in the photometric sensitivity over the instrument's lifetime. Monitoring the sensitivity is critical to providing the most accurate instrument calibration to the user community. The sensitivity change has been characterized in the past through staring mode observations of globular clusters \parencite{2022Bajaj}, scanning mode photometry of an open cluster \parencite{Som2021}, and grism observations of CALSPEC standard stars \parencite{Som2024}. 
\bigbreak
Non-crowded regions of galactic globular clusters are excellent targets to measure the sensitivity of the IR detector. If a cluster has been imaged multiple times with the same observational strategy (filter, exposure time, sample sequence, etc.) over long periods of time in different epochs, then changes in sensitivity can easily be measured by comparing the photometric measurements of individual stars across the epochs. \textcite{2022Bajaj} measured the sensitivity of the IR channel over time using three clusters: 47~Tucanae (47Tuc), Messier~4 (M4), and Omega Centauri ($\omega$ Cen) with data spanning from 2009--2022, 2016--2022, and 2015--2022, respectively. \textcite{2022Bajaj} found that the IR channel appears to be losing sensitivity at an average rate of $\approx0.13\pm0.02\%$ per year.
\bigbreak
\textcite{2024Marinelli} found that the detector has a cumulative sensitivity loss of $\sim$1--2$\%$ since installation in 2009, and that the sensitivity changes appear to be wavelength-dependent, with the largest changes in shorter-wavelength filters. An updated photometry reference table \texttt{IMPHTTAB} with updated inverse sensitivity corrections was delivered to \texttt{calwf3} in 2024 \parencite{2024calamida}. Efforts to monitor the IR channel sensitivity remain ongoing.
\bigbreak
In this report, we continue to measure the sensitivity of the IR channel using staring mode photometry of two globular clusters in two different filters: 47Tuc (F160W) and M4 (F110W and F160W). The most recent study on sensitivity changes measured with clusters includes data up to 2023 and is presented in \textcite{2024Marinelli}. We include the additional program shown in \textcite{2024Marinelli} (PID 17260) and add data from three additional calibration programs (PIDs 17363, 17683, and 17963) to recalculate the IR sensitivity loss and identify if there are any substantial changes in the last 2--3 years. These programs add 3--7 visits per target ranging from February 2023--April 2026.
\section*{Methods}
Our analysis procedure for the data from our three additional programs is adopted from \textcite{2022Bajaj} and is described in greater detail in that report. We provide a brief summary of the methods as follows:
\begin{enumerate}
    \item We run \texttt{hst1pass} \parencite{Anderson2022} on all calibrated \texttt{flt} exposures of our targets to obtain precise positions for all sources in the image. 
    \item We perform a two-step alignment using the DrizzlePac \parencite{2025dpacbook} function \texttt{Tweakreg} \parencite{2010Fruchter} with the \texttt{hst1pass} catalogs as inputs. First, images are aligned to Gaia DR2 using the methodology described in \textcite{2017Bajaj}. The second step then aligns all images of the same target together to minimize relative astrometric error. Individual catalogs are matched into a master catalog, containing all records of unique sources measured in the multiple exposures.
    \item Using the positions returned from the \texttt{hst1pass} catalogs, we perform aperture photometry with a target aperture radius of 3 pixels and sky annuli of 9--13 pixels.
    
    \item Sources within a certain brightness range and above a minimum number of detections across the sample of exposures for a given target are selected to avoid systematically biasing the sensitivity measurement. The selections differ for each target, and are described in further detail in the Results section.
\end{enumerate}
To mitigate the effects of persistence, we compute the sensitivity change twice: once using all images for a given target/filter combination (which would suffer from self-persistence), and another time using only the first image of every visit (which would only be affected by external persistence from other observations taken before the visit, if present). The second method trades larger number of measurements of sources for reduced systematic noise.
\bigbreak
We also compute the relative sensitivity in two ways. In the first method, we compute the median magnitude for each star across all exposures in which the star was detected. We then calculate the magnitude offset, which is defined to be the median magnitude of that star minus the individual magnitude measurement of the same star in each image. We then calculate the median magnitude offset of all stars in that image, and plot this value against the exposure date of each exposure. Finally, we perform a linear fit to these measurements to calculate the sensitivity change per year.
\bigbreak
In the second method, we compare the individual magnitude measurements versus exposure date of the same star across all exposures, and perform a linear fit to the measured magnitudes versus time. We then take the slope of the linear fit for each of the stars and calculate the median to estimate the change in sensitivity.
\bigbreak
The first method shows how the photometric sensitivity evolves between exposures, by characterizing relative photometric offsets of well measured stars.  However, due to sampling biases (not all stars are detected in every image, biasing the median magnitude), the offsets may be systematically shifted.  The second method is not affected by the bias in the median because the slope for an individual star is independent of the overall vertical shift.  However, as the fit model is a simple linear function, any nonlinear behavior is not accurately reflected by the slope. While the second method is still susceptible to some sampling bias, it is less susceptible than the first method because in the second method, a majority of the stars are detected in a majority of the images. 
\bigbreak
By combining the two methods above with the two sets of 1.) all exposures or 2.) only first exposures in each visit, we have four combinations:
\begin{enumerate}
    \item Sensitivity change measured with the \textit{median magnitude offsets} of stars in an image versus time for sources in \textit{every exposure}.
    \item Sensitivity change measured with the \textit{median magnitude offsets} of stars in an image versus time for sources in the \textit{first exposure of every visit}.
    \item Sensitivity change measured from the \textit{median slope} calculated from individual slope measurements of sources in \textit{every exposure}.
    \item Sensitivity change measured from the \textit{median slope} calculated from individual slope measurements of source in the \textit{first exposure of every visit}.
\end{enumerate}
\section*{Results}
We highlight our results for each target/filter combination individually below.
\subsection*{47Tuc (F160W)}
\begin{figure}[H]
\begin{center}
\includegraphics[width=\linewidth]{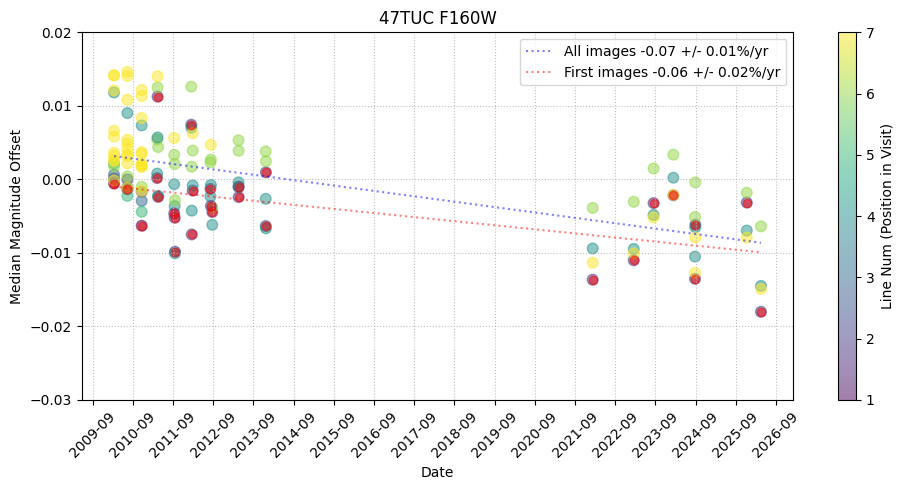}
\end{center}
\vspace{-1em}
\caption{\textit{Median magnitude offset for 47Tuc F160W images. The color of the points corresponds to the chronological position (\texttt{LINENUM}) of each exposure in each visit. Red points marks the first exposures in each visit. Images taken later in the same visit are more likely to be affected by self-persistence, especially in the early visits when persistence mitigation tactics were not implemented. A line is fitted to all exposures (blue dotted line) and the first exposure in every visit (red dotted line). The sensitivity loss (slope) and its uncertainties are displayed in the legend in units of \% per year.}
}
\label{fig:47tuc_scatter}
\end{figure}

Observing details of past 47Tuc observations and the star selection for 47Tuc exposures are described in \textcite{2022Bajaj}, but for completeness we repeat the information here: Observations of 47Tuc were taken in F160W from 2010--2013 as part of the IR linearity monitoring program. \textcite{2022Bajaj} adds an additional visit taken in 2022 to better constrain sensitivity change rates. This study adds seven additional visits ranging from February 2023 to April 2026, two of which are already included in \textcite{2024Marinelli}. Table \ref{tab:47Tuc} in the Appendix lists the visit name, proposal IDs, observed date, start time, and number of images \textit{N} per visit for all additional 47Tuc exposures in this study.
\bigbreak
Stars were selected if they appeared in at least 95 of the 132 47Tuc exposures and were brighter than ST$_{mag}$ $\leq$ 22.5. Additionally, we select well-fit stars with mean fit quality values of 0~$\leq$~\textit{q}~$\leq$~0.05 \parencite{Anderson2022}.
\bigbreak
Figure \ref{fig:47tuc_scatter} shows the median magnitude offset for 47Tuc exposures. A line is fit to the data with its slope equivalent to the rate of sensitivity loss in the IR channel using 1.) all exposures (blue dotted line) and 2.) first exposures in each individual visit (red dotted line). We find that the IR channel is losing 0.07\% $\pm$ 0.01\% per year from fitting every 47Tuc F160W exposure. We also find that the  latter result agrees with the 0.06\% $\pm$ 0.02\% per year from fitting using only the first exposures of each epoch, within their uncertainties. This sensitivity loss over time is the same as measured in \textcite{2024Marinelli} and is lower than what was initially measured in \textcite{2022Bajaj} (-0.12\% $\pm$ 0.02\% for all exposures and -0.10\% $\pm$ 0.04\% for the first exposure in every visit). The smaller measured sensitivity loss can be attributed to: 1.) the degradation in the sensitivity of the detector is plateauing over time and/or 2.) the observational strategy utilized in later visits reduces the self-persistence effect that biases the earlier data from 2009--2014.

\begin{figure}[H]
\begin{center}
\includegraphics[width=.95\linewidth]{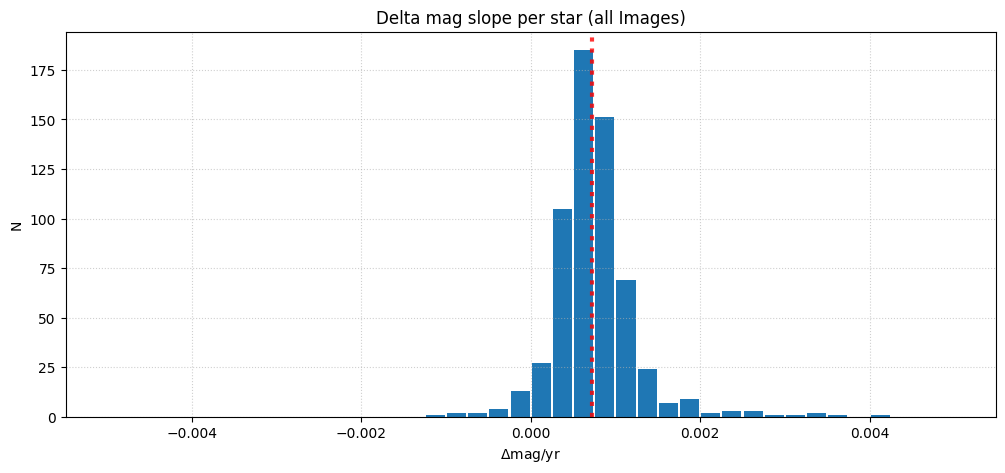}
\end{center}
\vspace{-1em}
\caption{\textit{Slopes for selected stars in all 95 47Tuc F160W exposures used. The median slope ($7\pm0.21\times 10^{-4}$ magnitudes per year) is shown as a red dotted line. Positive values indicate increasing magnitudes (lower measured flux).}
}
\label{fig:47tuc_allimghist}
\end{figure}

\begin{figure}[H]
\begin{center}
\includegraphics[width=.95\linewidth]{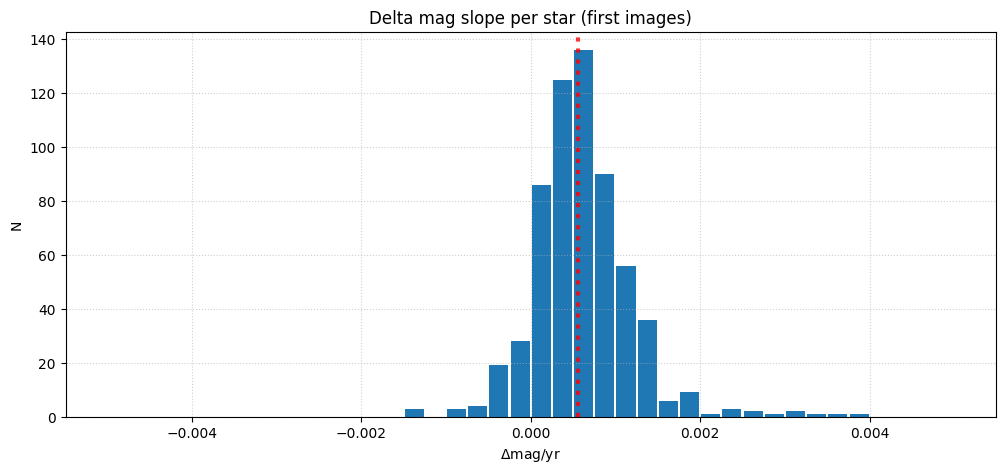}
\end{center}
\vspace{-1em}
\caption{\textit{Slopes for selected stars in the first exposures (27 total exposures) of every visit for F160W 47Tuc images. The median slope ($6\pm0.25\times 10^{-4}$ magnitudes per year) is shown as a red dotted line. }
}
\label{fig:47tuc_1stimghist}
\end{figure}

Figures \ref{fig:47tuc_allimghist} and \ref{fig:47tuc_1stimghist} show the sensitivity change measured from the median slope calculated from selected sources in 1.) every exposure and 2.) first exposure in each visit, respectively, for 47Tuc exposures. The median slope is plotted as a red dotted line. In both cases, the median slope agrees with what is measured in Figure \ref{fig:47tuc_scatter} within its uncertainties when using the same method (all exposures versus first exposure in each visit).   As the slopes in the second histogram are computed from fewer images, the errors on the individual slopes increase, thus widening the histogram.  The same behavior follows for the other targets.

\subsection*{M4 (F110W and F160W)}
Past M4 exposures were taken in 2012, 2017, 2021, and 2022 in F110W, and 2012, 2021, and 2022 in F160W as part of GO programs 12602, 14725, and CAL program 16512. Four additional visits per filter are added to this study, taken in August 2023, January 2025, April 2025 and February 2026. Note we did not observe M4 in Cycle 31 (2024). Exposures taken in 2012 used a 4-point dither of about three pixels between exposures, which was then repeated from the original starting position to obtain additional exposures. In later exposures (2021 onward), this dither is increased to $\sim$28 pixels between exposures to better mitigate persistence.  Tables \ref{tab:M4F110W} and \ref{tab:M4F160W} list the visit name, proposal IDs, observed date, start time, and \textit{N} number of images per visit for all new F110W and F160W M4 exposures, respectively.
\bigbreak
Stars were selected if they were detected in at least 15 of the 28 total exposures, had ST$_{mag}$ $\leq$ 21, and a mean fit quality value between 0 $\leq$ \textit{q} $\leq$ 0.06. These selections are applied to both the F110W and F160W exposures.
\subsubsection*{F110W}
\begin{figure}[H]
\begin{center}
\includegraphics[width=\linewidth]{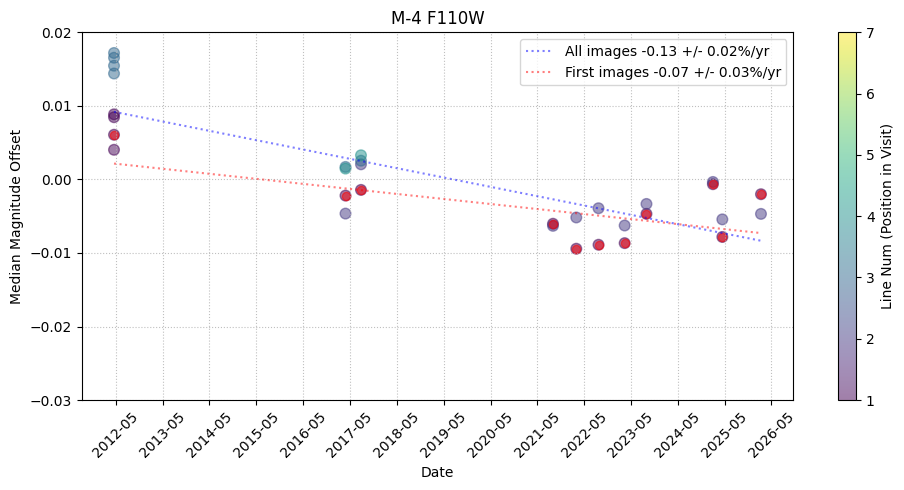}
\end{center}
\vspace{-1em}
\caption{\textit{Same as Figure \ref{fig:47tuc_scatter}, but for M4 exposures taken in F110W. M4 exposures taken in 2012 uses 4-point dither of about three pixels between exposures, however in later exposures this dither is increased to $\sim$28 pixels between exposures to mitigate persistence.}
}
\label{fig:M4_F110Wscatter}
\end{figure}

Figure \ref{fig:M4_F110Wscatter} shows the median magnitude offset for M4 exposures, with a linear fit overplotted  using 1.) all exposures (blue dotted line) and 2.) first exposures in each individual visit (red dotted line). Using F110W M4 exposures, we find that the IR channel is losing sensitivity at a rate of 0.13\% $\pm$ 0.02\% per year when using every exposure. We find a lower decrease in sensitivity rate of 0.07\% $\pm$ 0.03\% per year when using just the first exposures of every visit.

\begin{figure}[H]
\begin{center}
\includegraphics[width=.95\linewidth]{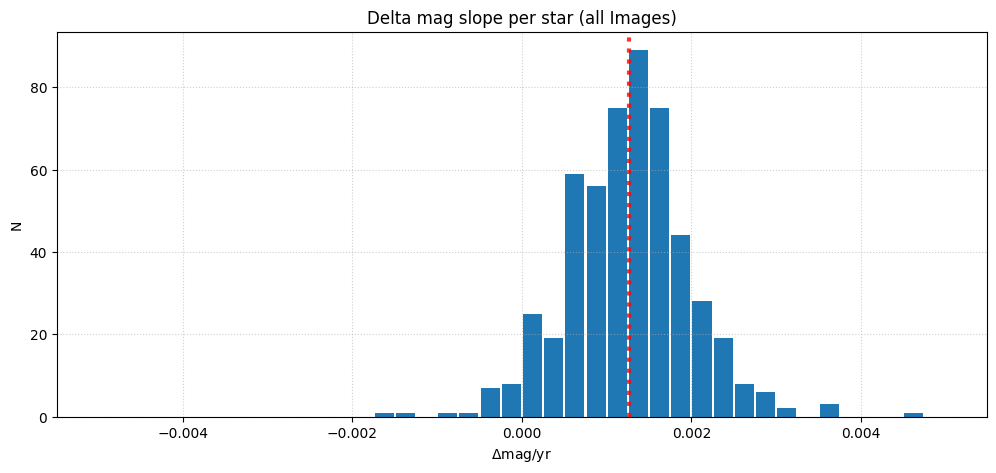}
\end{center}
\vspace{-1em}
\caption{\textit{Same as Figure \ref{fig:47tuc_allimghist}, but for all 15 F110W M4 exposures used. The median slope is calculated to be $12\pm0.34\times 10^{-4}$ magnitudes per year.}
}
\label{fig:M4_allimghist_f110w}
\end{figure}

\begin{figure}[H]
\begin{center}
\includegraphics[width=.95\linewidth]{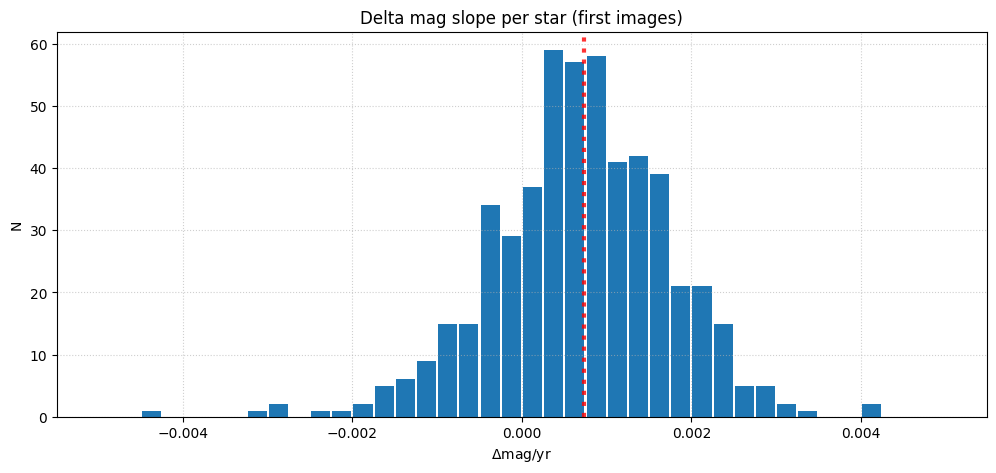}
\end{center}
\vspace{-1em}
\caption{\textit{Same as Figure \ref{fig:47tuc_1stimghist}, but for the first F110W M4 exposure in each visit (9 total exposures). The median slope is calculated to be $7\pm0.50\times 10^{-4}$ magnitudes per year.}
}
\label{fig:M4_1stimghist_f110w}
\end{figure}

Figures \ref{fig:M4_allimghist_f110w} and \ref{fig:M4_1stimghist_f110w} show the sensitivity change measured from the median slope calculated from selected sources in 1.) every exposure and 2.) first exposure in each visit, respectively, for M4 F110W exposures. The median slope in both cases agrees with the rate of sensitivity loss measured in Figure \ref{fig:M4_F110Wscatter} within uncertainties when using the same set of exposures (all exposures vs. first exposure in each visit).
 
\subsubsection*{F160W}
\begin{figure}[H]
\begin{center}
\includegraphics[width=\linewidth]{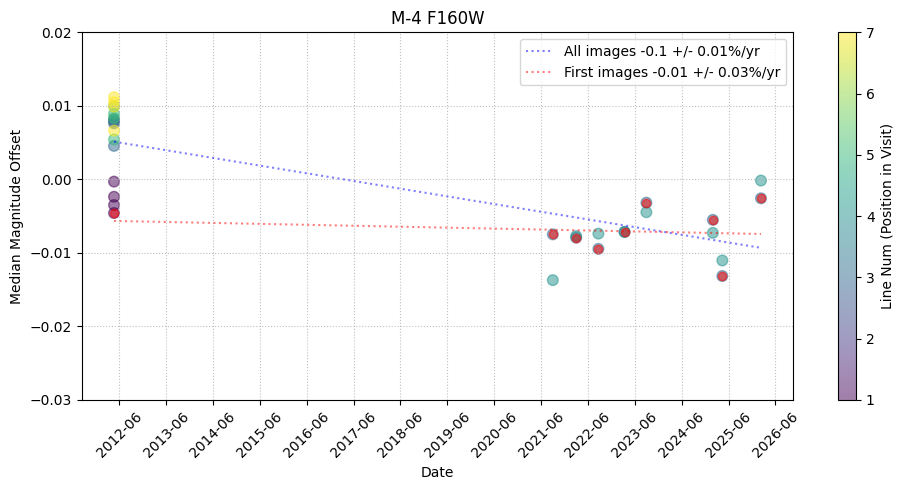}
\end{center}
\vspace{-1em}
\caption{\textit{Same as Figure \ref{fig:47tuc_scatter}, but for F160W M4 exposures. Early M4 exposures taken in 2012 are heavily affected by persistence, due to the use of repeated dither positions.}
}
\label{fig:M4_F160Wscatter}
\end{figure}

\begin{figure}[H]
\begin{center}
\includegraphics[width=.95\linewidth]{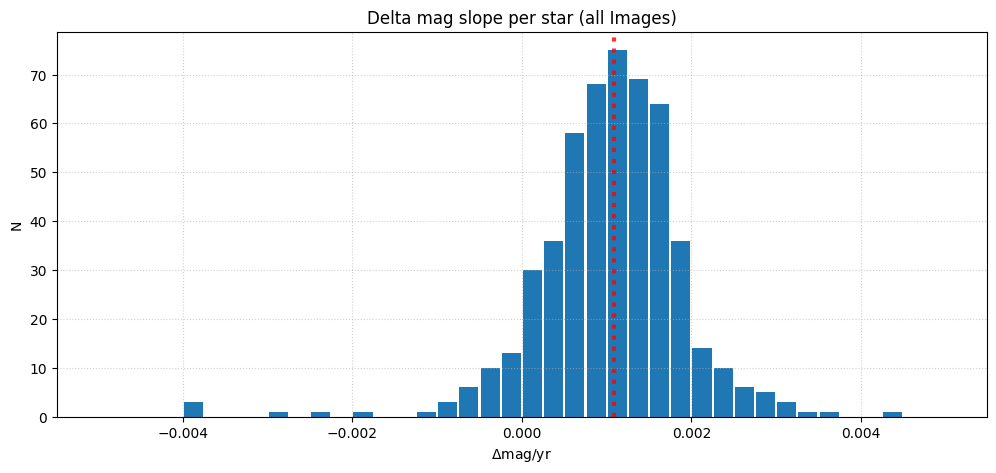}
\end{center}
\vspace{-1em}
\caption{\textit{Same as Figure \ref{fig:47tuc_allimghist}, but for F160W M4 exposures (15 total exposures used). The median slope is calculated to be $11\pm0.44\times 10^{-4}$ magnitudes per year.}
}
\label{fig:M4_allimghist_f160w}
\end{figure}

\begin{figure}[H]
\begin{center}
\includegraphics[width=.95\linewidth]{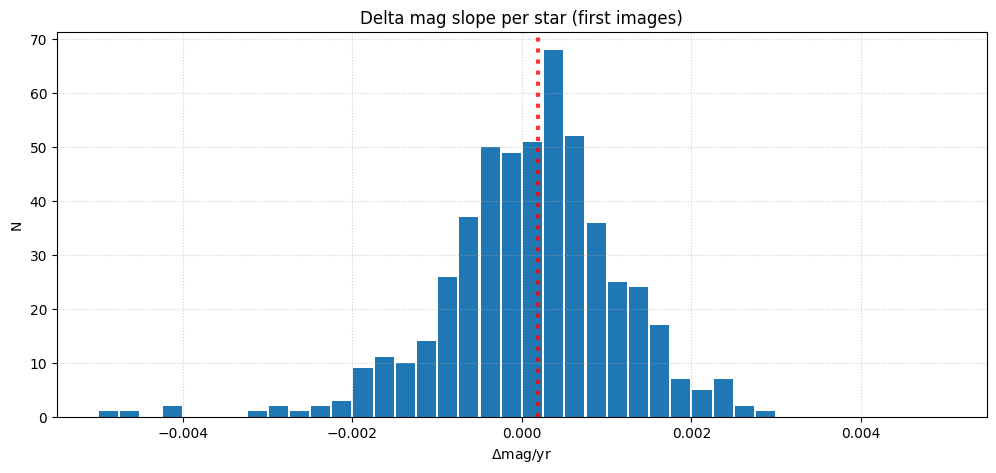}
\end{center}
\vspace{-1em}
\caption{\textit{Same as Figure \ref{fig:47tuc_1stimghist}, but for F160W M4 exposures (7 total exposures). The median slope is calculated to be $2\pm0.48\times 10^{-4}$ magnitudes per year.}}
\label{fig:M4_1stimghist_f160w}
\end{figure}

Figure \ref{fig:M4_F160Wscatter} shows the median magnitude offset for F160W M4 exposures, with the rate of sensitivity loss shown when measured using 1.) all F160W images and 2.) first F160W image in every visit. As mentioned in \textcite{2022Bajaj}, the early F160W exposures of M4 taken in 2012 are heavily affected by persistence, due to the use of the same dither positions \parencite{2019Bajaj}, so these results are less trustworthy. The rate of sensitivity loss measured using all images is 0.10\% $\pm$ 0.01\% per year. We find a much flatter decrease in sensitivity over time using the first exposure in every visit, with a rate of 0.01\% $\pm$ 0.03\% per year.
Figures \ref{fig:M4_allimghist_f160w} and \ref{fig:M4_1stimghist_f160w} show the sensitivity change measured from the median slope calculated from selected sources in 1.) every exposure and 2.) first exposure in each visit, respectively, for M4 F160W exposures. Similarly to the previous target/filter combinations, the median slope measured in both cases agrees with the sensitivity loss measured in Figure \ref{fig:M4_F160Wscatter} within uncertainties.

\section*{Conclusion}
We continue to measure the sensitivity of the WFC3/IR channel using staring mode photometry of the globular clusters 47Tuc and M4, with the addition of data from four monitoring programs in Cycles 30 -- 33. These programs added 3--7 visits per target ranging from February 2023 - April 2026. Table \ref{tab:avgtab} summarizes the sensitivity loss measured in this study for all relevant target–filter combinations.

\begin{table}[H]
\centering
\caption{\textit{Summary of the measured sensitivity loss for sampled stars observed in each of the targets exposure with the addition of new visits from three monitoring programs added to this study. }}
\label{tab:avgtab}
\begin{tabular}{|c|c|c|c|}
\toprule
Target & Filter & Median Slope (\%/year) & Median Slope (First Images) (\%/year) \\ 
\midrule
47Tuc & F160W & -0.07 $\pm$ 0.01 & -0.06 $\pm$ 0.02 \\
M4 & F160W & -0.10 $\pm$ 0.01 & -0.01 $\pm$ 0.03  \\
M4 & F110W & -0.13 $\pm$ 0.02 & -0.07 $\pm$ 0.03  \\
\bottomrule
\end{tabular}
\end{table}

\bigbreak
We find that in both cases (all exposures versus first exposure), the rate of sensitivity loss—measured after including additional visits from the three new monitoring programs—is lower for all relevant target/filter combinations than that reported by \textcite{2022Bajaj}. This may indicate that over time the rate at which the IR channel is losing sensitivity is decreasing. However, \textcite{2022Bajaj} had a much smaller baseline of exposures taken after 2019 that were strategically observed to avoid persistence. It is therefore plausible that the lower sensitivity loss rate measured in this report more accurately reflects the true long-term trends, whereas the higher rates reported previously may be overestimated due to strong persistence effects in early observations.
\bigbreak
\cite{2024Marinelli} adopted a time-dependent sensitivity correction value stored in the \texttt{IMPHTTAB} of 0.12\% $\pm$ 0.003\% per year and 0.06\% $\pm$ 0.005\% per year for F110W and F160W respectively. We find that the correction values from \cite{2024Marinelli} agree well with the sensitivity loss measured from cluster observations using only the first images in each epoch at redder wavelengths (F160W), but shows a discrepancy at bluer wavelengths (F110W). This difference is most likely driven by the limited constraint on early F110W photometry in our dataset, as only a single pre-2016 epoch is available for M4 observations taken in F110W, and would most likely resolve as more epochs are observed and the sensitivity loss in F110W becomes flatter. The \texttt{IMPHTTAB} reference file that contains the correction to the time-dependent sensitivity remains unchanged from the sensitivity measurements of this report.
\bigbreak
The WFC3/IR channel sensitivity continues to be measured every year with monitoring programs that observes star clusters such as the ones in this report. Current program 17963 of Cycle 33 is scheduled to observe an additional visit of 47Tuc in mid-2026. 
\section*{Acknowledgements}
The authors are especially grateful to Mitchell Revalski for his thorough review of this report, as well as Peter McCullough, Mariarosa Marinelli, Jennifer Mack, and Norman Grogin for their helpful reviews and feedback.
\newpage
\printbibliography
\newpage
\section*{Appendix}
\begin{table}[H]
\centering
\caption{\textit{Visit name, proposal IDs, observed date, start time, and number of images N in the visit for all additional F160W 47Tuc exposure added in this study (Cycles 30, 31, 32, and 33. One additional observation of 47 Tuc is planned for mid-2026 under Proposal 17693 (Cycle 33). For details on past observations, see \textcite{2022Bajaj}.}}
\label{tab:47Tuc}
\begin{tabular}{|c|c|c|c|c|}
\toprule
Visit & Proposal ID & Date-Obs & Start Time & N Images\\ 
\midrule
if2z02 & 17260 & 2023-02-16 & 03:47:32 & 4 \\
if2z03 & 17260 & 2023-08-14 & 22:56:00 & 4 \\
if5i01 & 17363 & 2024-08-28 & 18:39:05 & 4 \\
if5i02 & 17363 & 2024-02-11 & 02:28:53 & 4 \\
if5i03 & 17363 & 2024-08-26 & 01:55:19 & 4 \\
ifj5b1 & 17683 & 2025-12-09 & 03:48:42 & 4 \\
ifmua3 & 17963 & 2026-04-17 & 04:52:46 & 4 \\

\bottomrule
\end{tabular}
\end{table}

\begin{table}[H]
\centering
\caption{\textit{Same as Table \ref{tab:47Tuc}, but for M4 F110W exposures observed in Cycles 30, 32, and 33 (no observations of M4 were taken during Cycle 31).}}
\label{tab:M4F110W}
\begin{tabular}{|c|c|c|c|c|}
\toprule
Visit & Proposal ID & Date-Obs & Start Time & N Images\\ 
\midrule
if2z02 & 17260 & 2023-08-29 & 23:25:06 & 2 \\
ifj502 & 17683 & 2025-01-29 & 00:24:24 & 2 \\
ifj503 & 17683 & 2025-04-12 & 14:37:46 & 2 \\
ifmu02 & 17963 & 2026-02-07 & 23:20:36 & 2 \\

\bottomrule
\end{tabular}
\end{table}

\begin{table}[H]
\centering
\caption{\textit{Same as Table \ref{tab:47Tuc}, but for M4 F160W exposures observed in Cycles 30, 32, and 33 (no observations of M4 were taken during Cycle 31).}}
\label{tab:M4F160W}
\begin{tabular}{|c|c|c|c|c|}
\toprule
Visit & Proposal ID & Date-Obs & Start Time & N Images\\ 
\midrule
if2z01 & 17260 & 2023-08-29 & 23:49:06 & 2 \\
ifj502 & 17683 & 2025-01-29 & 00:48:24 & 2 \\
ifj503 & 17683 & 2025-04-12 & 15:01:46 & 2 \\
ifmu02 & 17963 & 2026-02-07 & 23:44:36 & 2 \\

\bottomrule
\end{tabular}
\end{table}

\end{document}